\documentclass{osa-article}

\journal{oe}
\articletype{Research Article}
\usepackage{multirow}
\providecommand{\keywords}[1]
{
  \small	
  \textbf{Keywords:} #1
}

\begin{document}

\title{Towards ultra-low-cost smartphone microscopy}

\author{Haoran Zhang,\authormark{1} Weiyi Zhang,\authormark{1} Zirui Zuo,\authormark{1} Jianlong Yang,\authormark{1,*}}

\address{\authormark{1}School of Biomedical Engineering, Shanghai Jiao Tong University, 1954 Huashan Road, Xuhui District, Shanghai 200030, China}

\email{\authormark{*}jyangoptics@gmail.com} %% email address is required



\begin{abstract*}
The outbreak of COVID-19 exposed the inadequacy of our technical tools for home health surveillance, and recent studies have shown the potential of smartphones as a universal optical microscopic imaging platform for such applications. However, most of them use laboratory-grade optomechanical components and transmitted illuminations to ensure focus tuning capability and imaging quality, which keeps the cost of the equipment high. Here we propose an ultra-low-cost solution for smartphone microscopy. To realize focus tunability, we designed a seesaw-like structure capable of converting large displacements on one side into small displacements on the other (reduced to $\sim$9.1\%), which leverages the intrinsic flexibility of 3D printing materials. We achieved a focus-tuning accuracy of $\sim5$ $\mu$m, which is 40 times higher than the machining accuracy of the 3D-printed lens holder itself. For microscopic imaging, we use an off-the-shelf smartphone camera lens as the objective and the built-in flashlight as the illumination. To compensate for the resulting image quality degradation, we developed a learning-based image enhancement method. We use the CycleGAN architecture to establish the mapping from smartphone microscope images to benchtop microscope images without pairing. We verified the imaging performance on different biomedical samples. Except for the smartphone, we kept the full costs of the device under 4 USD. We think these efforts to lower the costs of smartphone microscopes will benefit their applications in various scenarios, such as point-of-care testing, on-site diagnosis, and home health surveillance.
\end{abstract*}\\

\\

\textbf{Research highlights:} We propose a solution for ultra-low-cost smartphone microscopy. Utilizing the flexibility of 3D-printed material, we can achieve focusing accuracy of $\sim5$ $\mu$m. Such a low-cost device will benefit point-of-care diagnosis and home health surveillance.
\begin{figure}[h]
\centering
\centering\includegraphics[width=0.5\linewidth]{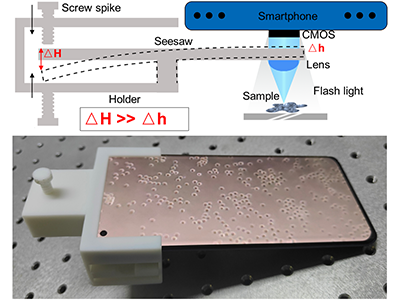}
\caption{Graphical abstract image}
\label{Graphical abstract image}
\end{figure}
%%%%%%%%%%%%%%%%%%%%%%%%%%  body  %%%%%%%%%%%%%%%%%%%%%%%%%%
\section{Introduction}
Smartphones have become an indispensable tool in our daily lives, and their functions are no longer limited to communication. With the development of semiconductors, sensors, artificial intelligence, etc., smartphones have the ability to acquire, process, and exchange information comparable to that of laptops and even desktop computers. At the same time, the advantage of portability makes them more frequently used. On the other hand, the rapid development of miniaturized optical imaging components for smartphones, such as lenses, CMOS sensors, and stabilization, has made them a desirable and universal platform for microscopy. Numerous research attempts have emerged in recent years to develop smartphone-based biomedical instrumentation for in vitro and in vivo diagnostics, noninvasive monitoring, and therapeutic guidance (see \cite{banik2021recent,hussain2021smartphone,hunt2021smartphone} and references therein).\\
\indent The outbreak of COVID-19 has caused a global shortage of medical resources. Over the past three years, queuing up for the nucleic acid test of COVID-19 has become a routine for us, consuming a lot of time, money, and human resources. Although antigen testing enables home self-testing, it is not accurate enough to provide diagnostic conclusions. Smartphones have attracted the attention of many researchers due to their popularity and portability, and have initially demonstrated their potential for virus detection \cite{ravindran2021smartphone}. For example, Fozouni \textit{et al.} reported the development of amplification-free detection of SARS-CoV-2 with CRISPR-Cas13a and mobile phone microscopy \cite{fozouni2021amplification}. Ganguli \textit{et al.} demonstrated isothermal RT-LAMP nucleic acid-based detection of SARS-CoV-2 with a smartphone-based instrument \cite{ganguli2020rapid}. \\
\indent However, most smartphone-based biomedical detection technologies remain in the stage of laboratory demonstration, with few commercial applications promoted \cite{hunt2021smartphone}, one of the main reasons is that the manufacturing costs of such devices are relatively high by household standards. The imaging performance of optical microscopy can be characterized as \cite{jin2020deep}
\begin{equation}
    d \propto \frac{\lambda}{NA^2} \propto \frac{\Delta r^2}{\lambda},
    \label{eq0}
\end{equation}
where $d$ is the depth of focus, $\lambda$ is the central wavelength of illumination, $NA$ is the numerical aperture of objective lens, and $\Delta r$ is the optical resolution. For micron and submicron optical resolutions, the depth of focus ranges from a few microns to tens of microns.\\
\indent To achieve optical focusing at such precision, a straightforward solution is to use laboratory-grade optomechanical components. For example, Sun \textit{et al.} constructed a smartphone-based dark-field microscope for quantitative nanoparticle analysis. Although they claimed that this is a low-cost device, more than 1,000 USD was spent to fabricate it \cite{sun2018low}. Several research groups have used 3D-printed optomechanical components to reduce the costs \cite{knowlton20173d,orth2018dual,liu2021pocket}, but they employed fixed-focus microscopic imaging (with non-adjustable working distance), which affects the compatibility of such devices. Besides, due to the popular fused deposition molding 3D printing has a processing accuracy of $\sim200$ $\mu$m, the 3D-printed optomechanical components usually require additional processing to obtain accurate dimensions or need to be selected from multiple prints. On the other hand, to ensure imaging quality, many smartphone microscopes employed the objective lenses of tabletop microscopes and transmitted illumination setups \cite{zhu2011optofluidic,ganguli2017hands,koydemir2015rapid,sun2018low,trofymchuk2021addressable,lee2021smartphone}, which further increases their costs and size.\\
\indent To address the above issues, many optical innovations have been developed. Lee \textit{et al.} \cite{lee2014fabricating} and Szydlowski \textit{et al.} \cite{szydlowski2020cell} used droplets of different materials as the objective lens for smartphone microscopy. Dai \textit{et al.} further added color dyes to the droplet lenses for fluorescence imaging \cite{dai2019colour}. Although this approach can significantly reduce device costs and size, its imaging quality is poor due to the difficulty of controlling the surface precision of the droplet lenses and their uneven spectral transparency. Similarly, off-the-shelf smartphone accessories including imprinted lenses have been developed \cite{satzuma,apexel} and they have better optical quality for microscopic imaging. However, these endeavors to lower the costs of lenses do not enable focus tunability. Song \textit{et al.} proposed a focus tuning method, which is
based on the dynamic deformation of an elastomeric membrane driven by adjustable hydraulic pressure \cite{song2020smartphone}. This method is simple and effective but has difficulties in aberration correction.\\
\indent In this paper, we propose an ultra-low-cost solution for smartphone microscopy, which tries to address the trade-off between the costs and imaging capabilities. The main contributions of our work include:
\begin{itemize}
    \item To realize focus tunability, we designed a seesaw-like structure capable of converting large displacements on one side into small displacements on the other, which takes advantage of the intrinsic flexibility of 3D printing materials. Our method achieved a focus-tuning accuracy of $\sim5$ $\mu$m, which is 40 times higher than the machining accuracy of the 3D printed lens holder itself.
    \item To compensate for the image quality degradation due to the use of an off-the-shelf smartphone camera lens (as the objective) and the built-in flashlight (as the illumination), we developed a learning-based image enhancement method. We leverage the cycle-consistency generative adversarial network architecture \cite{zhu2017unpaired} to establish the mapping from smartphone microscope images to benchtop microscope images, which do not need paired field of view (FOV) data for training.
    \item We kept the full costs of our device under 4 USD (except for the smartphone), which is, to the best of our knowledge, the cheapest smartphone microscope reported. We think such a low-cost device will benefit many healthcare scenarios, such as point-of-care testing, on-site diagnosis, and home health surveillance.
\end{itemize} 
\indent We arrange this paper as follows: We describe the proposed focus tuning and image enhancement methods in Section 2. In Section 3, we give the results that demonstrate the capabilities of our methods, including focus tunability, imaging quality, and imaging of various biomedical specimens. We discuss the prospects and limitations of such methods in Section 4 and draw our conclusions in Section 5. 

\section{Methods and materials}
\subsection{Seesaw-like structure for focus tuning}
\begin{figure}[b]
\centering\includegraphics[width=\linewidth]{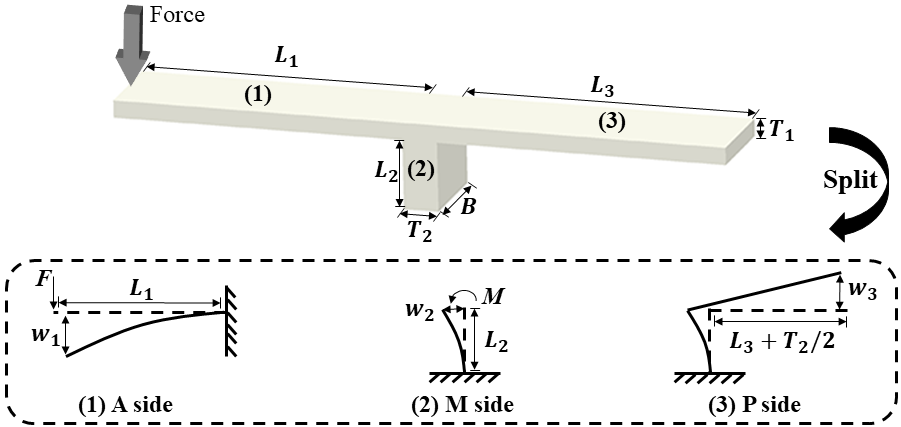}
\caption{Geometrical characteristics of the seesaw-like structure and its mechanical simplification for modeling. We split the seesaw-like structure into three parts, including (1) active (A) side, (2) middle (M) side, and (3) passive (P) side.}
\label{fig2}
\end{figure}

Our focus tuning method was inspired by the law of the lever, first articulated by the ancient Greek scientist Archimedes \cite{chondros2010archimedes}. It explains that when using a lever to lift or move an object, you can reduce the effort required by positioning the pivot point closer to the heavy load while applying your force farther away from the pivot point. Seesaw is a special type of lever in which the pivot point is located at the center, so the force on the object side is the same magnitude and opposite direction to the end that applies the force. by fixing the pivot point, we designed a seesaw-like structure capable of converting large displacements on one side of it into small displacements on the other side. Their displacement ratio is dependent on the intrinsic flexibility of materials. \\
\indent Figure~\ref{fig2} illustrates the geometrical characteristics of the seesaw-like structure and its mechanical simplification for modeling. The seesaw-like structure has a horizontal hanging beam and a vertical supporting beam, which are connected rigidly. We simplified the problem by splitting the seesaw-like structure into three parts. We named the force end (1) of the hanging beam as the active (A) side, the supporting beam (2) as the middle (M) side, and the free end (3) of the hanging beam as the passive (P) side. We then calculated the deflection and rotation angle of each side to achieve the displacement ratio between the A and P sides.\\
\indent We assumed the axis of the hanging beam is a continuous smooth curve after bending within the range of elastic strength, the deflection curve as a function of curvature $\rho$ can be written as \cite{gross2016mechanics}\\
\begin{equation}
\label{eq1}
    \frac{1}{\rho }\text{=}\frac{M}{E\cdot I},
\end{equation}
where \textit{M} is torque, \textit{E} is Young's modulus (or modulus of elasticity), and \textit{I} is the moment of inertia. We assumed the beams have uniform cross-sections. The rotation angle equation \textit{$\theta$}(\textit{x}) and deflection equation \textit{w}(\textit{x}) are obtained by the indefinite integral of horizontal position \textit{x}:\\
\begin{equation}
    \theta \left( x \right)=\frac{dw}{dx}=\int_{l}{\frac{M(x)}{E\cdot I}dx+C},
\label{eq2}
\end{equation}
\begin{equation}
    w\left( x \right)=\int_{l}{\left[ \int_{l}{\frac{M(x)}{E\cdot I}dx} \right]}dx+C\cdot x+D,
\label{eq3}
\end{equation}
where \textit{C} and \textit{D} are integration constants. The A side was treated as a cantilever beam model with a concentrated load acting on the end of the beam. By setting the torque and rotation angle at the rigid joint as 0, the maximum deflection and corresponding rotation angle at the A side can be derived:\\
\begin{equation}
    {{w}_{1}}=\frac{F\cdot {{L}_{1}}^{3}}{3E\cdot {I}_{1}} \quad and \quad {{\theta }_{1}}\text{=}\frac{F\cdot {{L}_{1}}^{2}}{2E\cdot {I}_{1}},
\end{equation}
where ${I}_{1}=\frac{B\cdot T_{1}^{3}}{12}$. We treated the M side as a cantilever beam model with torque acting on the joint, the maximum deflection and corresponding rotation angle can be written as\\
\begin{equation}
    {{w}_{2}}=\frac{M\cdot{{L}_{2}}^{2}}{2E\cdot{{I}_{2}}} \quad and \quad {{\theta }_{2}}\text{=}\frac{M\cdot {{L}_{2}}}{E\cdot {{I}_{2}}},
\end{equation}
where ${I}_{2}=\frac{B\cdot T_{2}^{3}}{12}$. The P side was supposed to be perpendicular to the M side at the joint. We ignored the deformation caused by gravity because of its small size. Thus the maximum deflection and corresponding rotation angle at the P side are determined by\\
\begin{equation}
    {{w}_{3}}=({{L}_{3}}+\frac{{{T}_{2}}}{2})\sin {{\theta }_{3}}\approx ({{L}_{3}}+\frac{{T}_{2}}{2}){{\theta }_{3}} \quad and \quad  {{\theta }_{3}}={{\theta }_{2}},
\end{equation}
where sin \textit{$\theta_{3}$} is replaced by $\theta_{3}$ because the rotation angle is very small. Given that \textit{$M=F\cdot L_{1}$}, we can deduce the vertical A:P displacement ratio:\\
\begin{equation}
\text{A:P}=\frac{{{w}_{1}}}{{{w}_{3}}}=\frac{{{I}_{2}}\cdot{{L}_{1}}^{2}}{3{{I}_{1}}\cdot{{L}_{2}}({{L}_{3}}+\frac{{{T}_{2}}}{2})}=\frac{T_{2}^{3}\cdot{{L}_{1}}^{2}}{3T_{1}^{3}\cdot{{L}_{2}}({{L}_{3}}+\frac{{{T}_{2}}}{2})},
\label{eq7}
\end{equation}
\indent We can see that the displacement ratio is not affected by the material property (\textit{i.e.,} Young's modulus \textit{E}). For a specific seesaw-like structure, the displacement ratio is constant, namely, the displacement of the P side changes linearly in proportion to the variation of the displacement of the A side, which is desirable in this focus tuning application. 
\subsection{Unpaired deep learning for image enhancement}
\begin{figure}[h!]
    \centering
    \includegraphics[width=\linewidth]{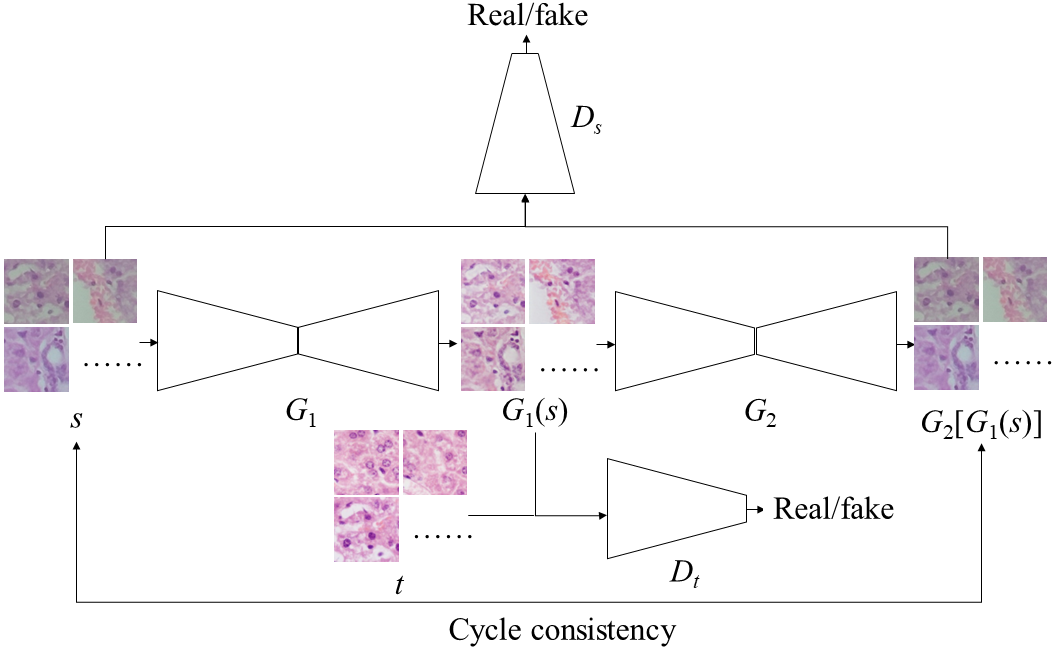}
    \caption{Architecture of our unpaired deep learning strategy for enhancing the image quality of smartphone microscopy.}
    \label{fig:network}
\end{figure}
Deep neural networks have universal approximation capability \cite{lu2020universal}, thus can be used to learn the mapping from low-quality microscopic images to high-quality ones. Rivenson \textit{et al.} used fully supervised deep learning to convert the images captured with a smartphone microscope to the images captured with a benchtop microscope \cite{rivenson2018deep,kim2023portable}. However, such methods require paired data for training deep-learning models. By paired data, we mean that images of the same FOV of the same sample were first obtained using a smartphone microscope and a benchtop microscope, respectively. The content in the two images is then corresponded at the pixel level using an alignment algorithm \cite{rivenson2018deep}. This process is time-consuming and laborious. Here we propose to use unpaired image enhancement deep learning to avoid the data pairing process. The unpaired deep learning only requires data from both smartphone and benchtop microscopes but does not need the FOV correspondence. We adopt the cycle-consistency generative adversarial network method \cite{zhu2017unpaired} for this purpose.\\
\indent Figure~\ref{fig:network} illustrates the architecture of our unpaired deep learning strategy. The cycle consistency consists of both the forward cycle and the backward cycle. We only illustrate the forward cycle here for simplicity. It enforces a transition cycle:
\begin{equation}
  s\rightarrow G_1(s) \rightarrow G_2[G_1(s)] \approx s.  
\end{equation}
The backward cycle has the same architecture but enforces an inverse transition cycle: 
\begin{equation}
t\rightarrow G_2(t) \rightarrow G_1[G_2(t)] \approx t,    
\end{equation} 
where $s$ and $t$ are the smartphone and benchtop microscopic images, respectively. $G_1$ and $G_2$ are the generators in Fig.~\ref{fig:network}. The Euclidean norm measures the distance between the input $s$ and the generated $s$. The discriminator aims to distinguish between generated samples $G_1(s)$ and real samples $t$. The training of this network is to solve:
\begin{equation}
    G_1^*,G_2^* = \arg\min_{G_1,G_2}\max_{D_s,D_t} \mathcal{L}(G_1, G_2, D_s, D_t),
\end{equation}
where $\mathcal{L}$ is the full objective including both the cycle consistency and adversarial discriminator introduced above. $D_s$ and $D_t$ are the discriminators to distinguish $G_1(s)$ and $t$, and $G_2(t)$ and $s$, respectively.
\subsection{Experimental setups}
\begin{figure}[h!]
    \centering
    \includegraphics[width=\linewidth]{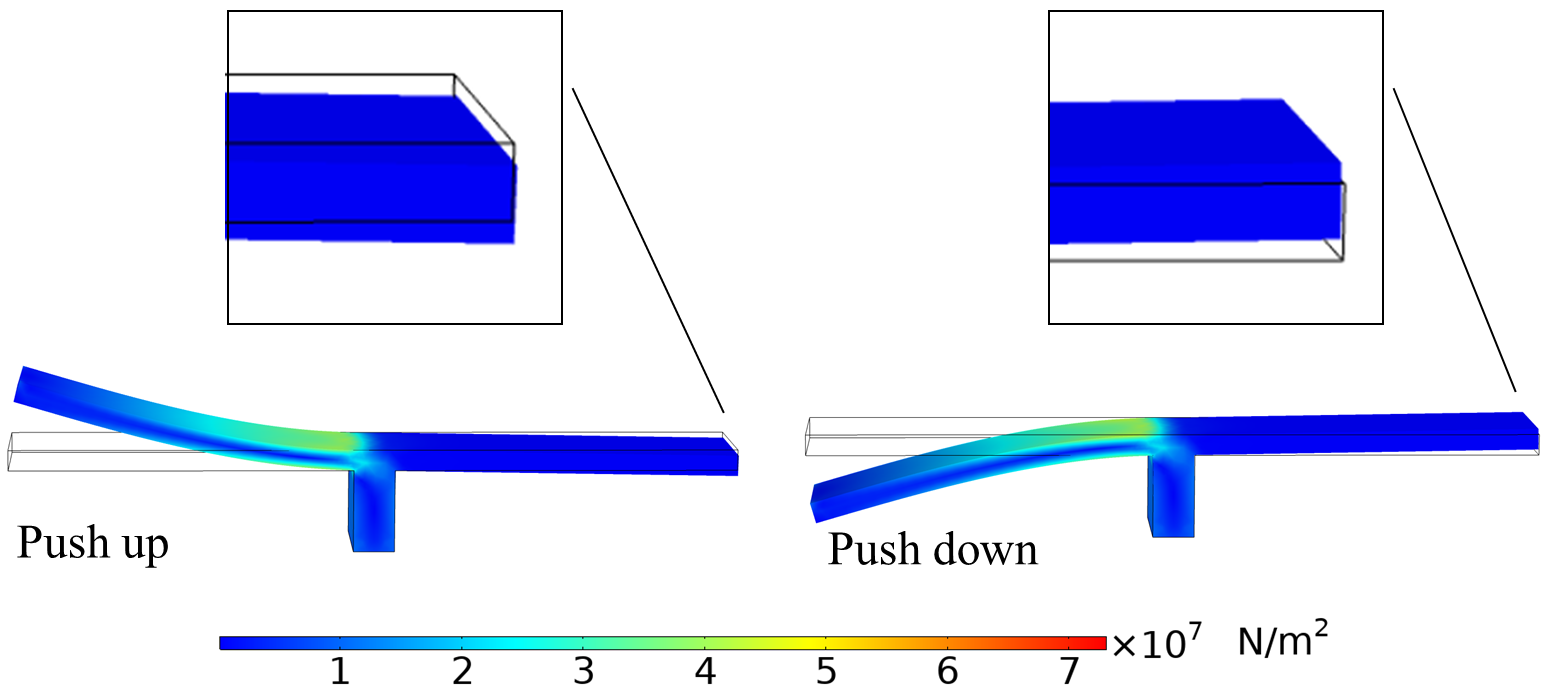}
    \caption{Numerical simulation examples of pushing up and down on the left (A) side. The zoom-in views show the induced displacement on the right (P) side. The color bar gives the magnitude of the applied force.}
    \label{fig:numerical}
\end{figure}
We conduct a numerical simulation of the seesaw-like structure for focus tuning using COMSOL Multiphysics software (version 5.4), to fully characterize its performance in parameter space and validate the theoretical model. Figure~\ref{fig:numerical}(a) gives two simulation examples of pushing up and down on the left (A) side.The zoom-in views show the induced displacement on the right (P) side. The color bar gives the magnitude of the applied force. We employ resin and nylon, two commonly used materials in 3D printing, in our numerical simulation. Their mechanical properties are listed in Table~\ref{tab:material}. Table~\ref{tab:geo} lists the geometric parameters of the seesaw-like structure used in this work.\\
\begin{table}[h!]
    \centering
        \caption{Mechanical properties of 3D printing materials used in our numerical simulation.}
    \begin{tabular}{c|ccc}
    \hline
        Material type & Young’s modulus (MPa) & Bending strength (MPa) & Density (kg/m$^3$)\\
        \hline
        Resin & 2700 & 73 & 1170 \\
        Nylon & 1300 & 46 & 1020 \\
        \hline
    \end{tabular}
    \label{tab:material}
\end{table}

\begin{table}
    \centering
        \caption{Geometric parameters of the seesaw-like structure used in this work.}
    \begin{tabular}{cccccc}
    \hline
       $L_1$ (mm) & $L_2$ (mm) & $L_3$ (mm) & $T_1$ (mm) & $T_2$ (mm) & B (mm)\\
       \hline
        25 & 6 & 25 & 3 & 1.5 & 8 \\
        \hline
    \end{tabular}
    \label{tab:geo}
\end{table}
\begin{figure}[h!]
    \centering
    \includegraphics[width=\linewidth]{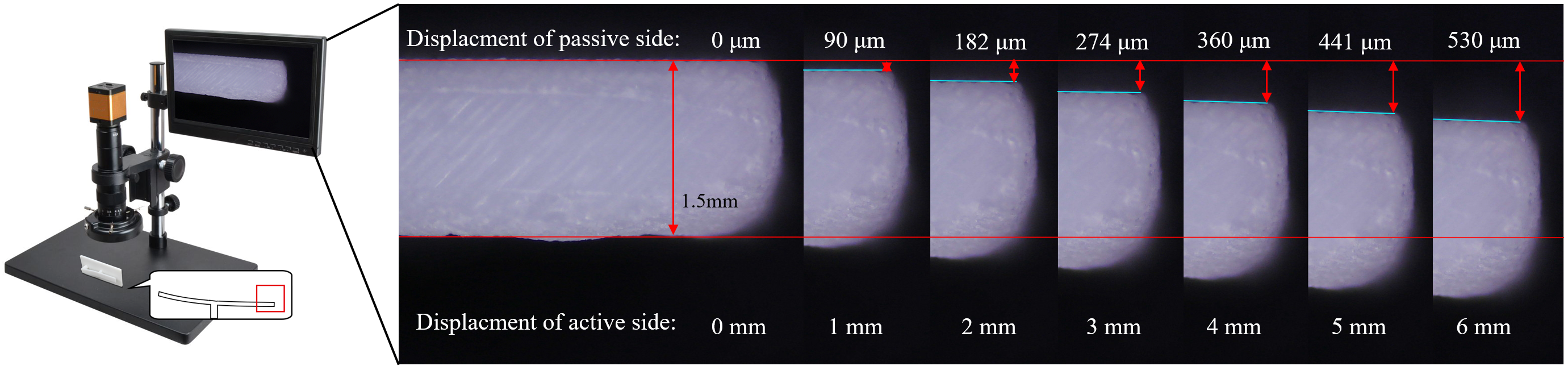}
    \caption{Experimental setup for measuring the micro-scale displacements of the passive (P) side.}
    \label{fig:ccd}
\end{figure}
\indent We employed a commercial microscope (GP-660V, Gaopin, China) with an optical magnification of 2$\times$ and a working distance of 95 mm to observe the micro-scale displacement of the passive (P) side. As demonstrated in Fig.~\ref{fig:ccd}, we can conduct a quantitative measurement of the displacement as the applied force of the active (A) side changes. \\
% In Equation \ref{eq7}, we found that the material property has no influence on the displacement ratio of the A and P sides. We have confirmed this finding in our numerical simulation, by comparing the results of using resin and nylon, which are two commonly-used FDM 3D printing materials. Their geometrical and material parameters used in the model validation are listed in Table~\ref{table1}. Because resin has a relatively larger Young's modulus, bending strength, and density than those of nylon, we employed it in the theoretical analysis and experimental investigation below. \\
\begin{figure}[h!]
\centering\includegraphics[width=\linewidth]{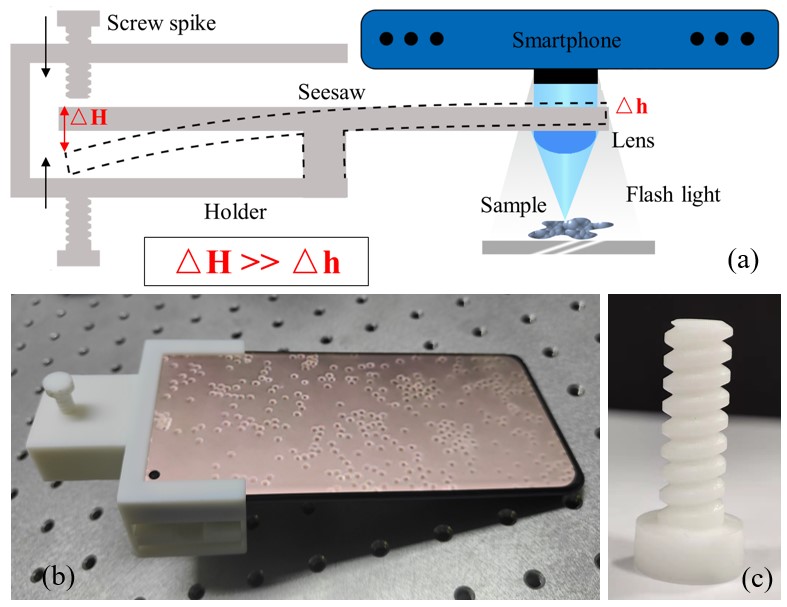}
\caption{Setups of our ultra-low-cost smartphone microscope. (a) Illustration of the imaging optics using the seesaw-like structure for focus tuning. (b) Photograph of the microscope. (c) Photograph of a 3D-printed screw spike.}
\label{fig5}
\end{figure}
\indent Figure~\ref{fig5} is the setups of our ultra-low-cost smartphone microscope. (a) illustrates the imaging optics using the seesaw-like structure for focus tuning. (b) is a photograph of the microscope. (c) is a photograph of a 3D-printed screw spike. We employ a Xiaomi 11 smartphone and its built-in camera with an aperture of \textit{f}/2.4. An off-the-shelf smartphone lens (the camera lens of iPhone 6s) is mounted on the seesaw-like structure to constitute a 4-f system with the built-in camera lens, which provides better imaging quality than the ball-lens- and droplet-based microscopes \cite{switz2014low}. We use the built-in flashlight to provide reflective illumination. The 3D-printed screw spike is used to apply force on the seesaw-like structure. Limited by the processing accuracy of 3D printing ($\sim200$ $\mu$m), it is difficult to print high-quality screw spikes with small diameters and thread pitches. So we design a screw spike with a diameter of 6 mm and a thread pitch of 2 mm.\\
\indent We use a personal computer with an NVIDIA GTX 1080 Ti GPU (12-GB onboard memory) for training the unpaired image enhancement model. We employ the open-source Pytorch implementation \cite{pytorch} of the cycle-consistent generative adversarial network \cite{zhu2017unpaired}. We resize input patches of the microscopic images to $256\times256$ pixels and select ResNet-based generator (9 residual blocks), which achieved better performance than U-Net in this task. The generator and discriminator were trained from scratch using the Adam optimizer ($\beta_1$ = 0.5 and $\beta_2$ = 0.999) with an initial learning rate of $2\times10^{-4}$ and a batch size of 1. We use 560 images each of smartphone microscopic images and benchtop microscopic (40x objective) images for model training. We capture 30 smartphone microscopic images from independent samples for testing, and benchtop microscopic images in the same FOV as ground truth for method evaluation. In the training of the GAN model, the loss curve converged after 180 epochs, so we stopped at 200 epochs to minimize the loss. Each training takes about 4 hours.
%%%%%%%%%%%%%%%%%%%%%3. Experiments and results%%%%%%%%%%%%%%%%%%%%%%%%%%%%%%%
\section{Results}
\subsection{Characterization of seesaw-like structure}
\begin{figure}[h!]
    \centering
    \includegraphics[width=\linewidth]{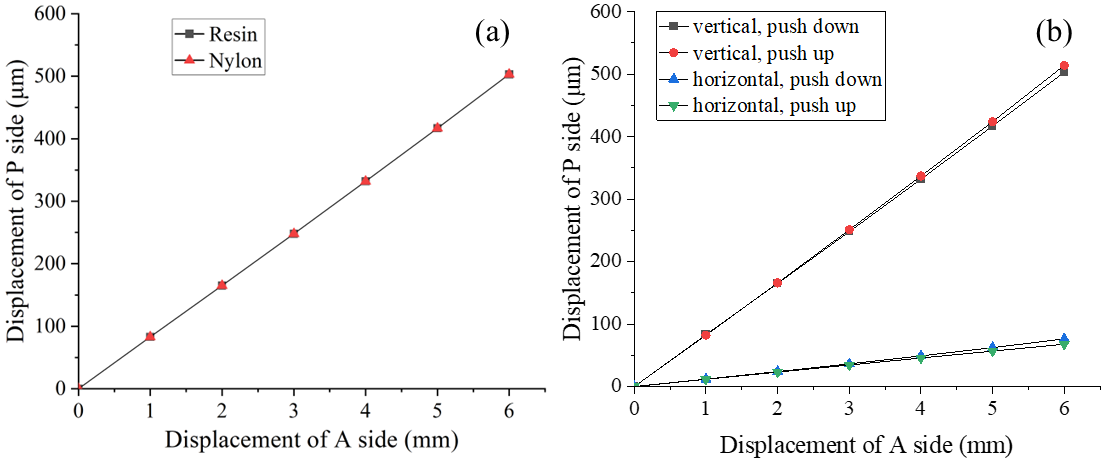}
    \caption{Numerical characterization of the seesaw-like structure. The displacement of the P side as a function of that of the A side (a) when two common 3D printing materials are used, and (b) when up and down forces are applied and the results of horizontal direction are also shown.}
    \label{fig:numResult}
\end{figure}
\begin{table}[h!]
\centering
\caption{Displacements of the P side ($\mu$m) and the A:P ratio.}
\begin{tabular}{cccccccc}
\hline
\multirow{2}{*}{} & \multicolumn{6}{c}{The given displacement of the A side}           & \multirow{2}{*}{A:P ratio} \\ \cline{2-7}
                     & 1 mm     & 2 mm      & 3 mm      & 4 mm      & 5 mm      & 6 mm      &                            \\ \hline
Theory               & 85      & 170      & 255      & 340      & 424      & 509      & 11.78                     \\
Simulation           & 82      & 166      & 251      & 337      & 424      & 514      & 11.84                      \\
Experiment                 & 90$\pm2.3$ & 182$\pm4.9$ & 274$\pm4.9$ & 360$\pm5.4$ & 441$\pm1.9$ & 530$\pm1.9$ & 11.19$\pm0.11$               \\ \hline
\end{tabular}
\label{table2}
\end{table}
\begin{figure}[h!]
    \centering
    \includegraphics[width=\linewidth]{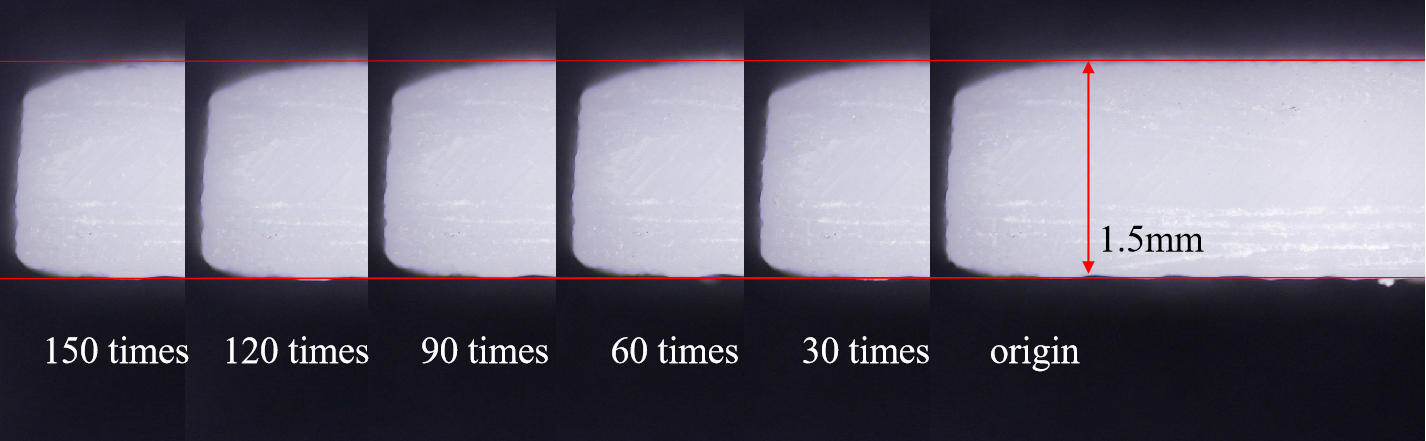}
    \caption{Repeatability test results of the seesaw-like structure.}
    \label{fig:repeat}
\end{figure}
Figure~\ref{fig:numResult} is a numerical characterization of the seesaw-like structure. We employ the material and geometric parameters listed in Table~\ref{tab:material} and \ref{tab:geo} in the simulation. Figure~\ref{fig:numResult}(a) is the displacement of the P side as a function of that of the A side, when two common 3D printing materials, resin and nylon, are used. The results show that the displacement ratio is not related to the choice of 3D printing materials, which is consistent with the prediction of the deduced theoretical model in Eq.~\ref{eq7}. Because resin has relatively larger Young's modulus, bending strength, and density than those of nylon, we employ it in the fabrication of our ultra-low-cost smartphone microscope as shown in Fig.~\ref{fig5}(b) and (c). Figure~\ref{fig:numResult}(b) is the displacement results when up and down forces are applied and the horizontal displacement is also included. We can see that both the push-up and push-down force on the A side lead to linear and consistent displacement alteration. Besides the vertical displacement, there is a horizontal displacement during the bending of the P side. Fortunately, this horizontal displacement is $\sim7.3$ times smaller than the vertical displacement, which has a neglectable influence on the FOV of our smartphone microscope as demonstrated in the experimental results below. \\
\indent Table~\ref{table2} lists the displacements of the P side ($\mu$m) as the given displacement of the A side increases, and the A:P ratio. We compare the theoretical prediction, numerical simulation, and experimental measurement. These results show that our theoretical model is consistent with the numerical simulation and experiments. The A:P ratios indicate that we achieve an improvement in focus tuning precision of $\sim11$ times from this setup of the seesaw-like structure.\\
\indent We also investigate the robustness of the 3D-printed seesaw-like structure in long-term usage. We repeatedly applied force to the A side of the seesaw-like structure 150 times. The force was controlled within the maximum bending strength. Figure~\ref{fig:repeat} is the microscopic recordings of the recovered A side after removing the force. We can see the hanging beam is able to completely return to its initial position after each bending, which enables the long-term utilization of this focus tuning method.
\subsection{Focus tunability using seesaw-like structure}
\begin{figure}[h!]
\centering\includegraphics[width=\linewidth]{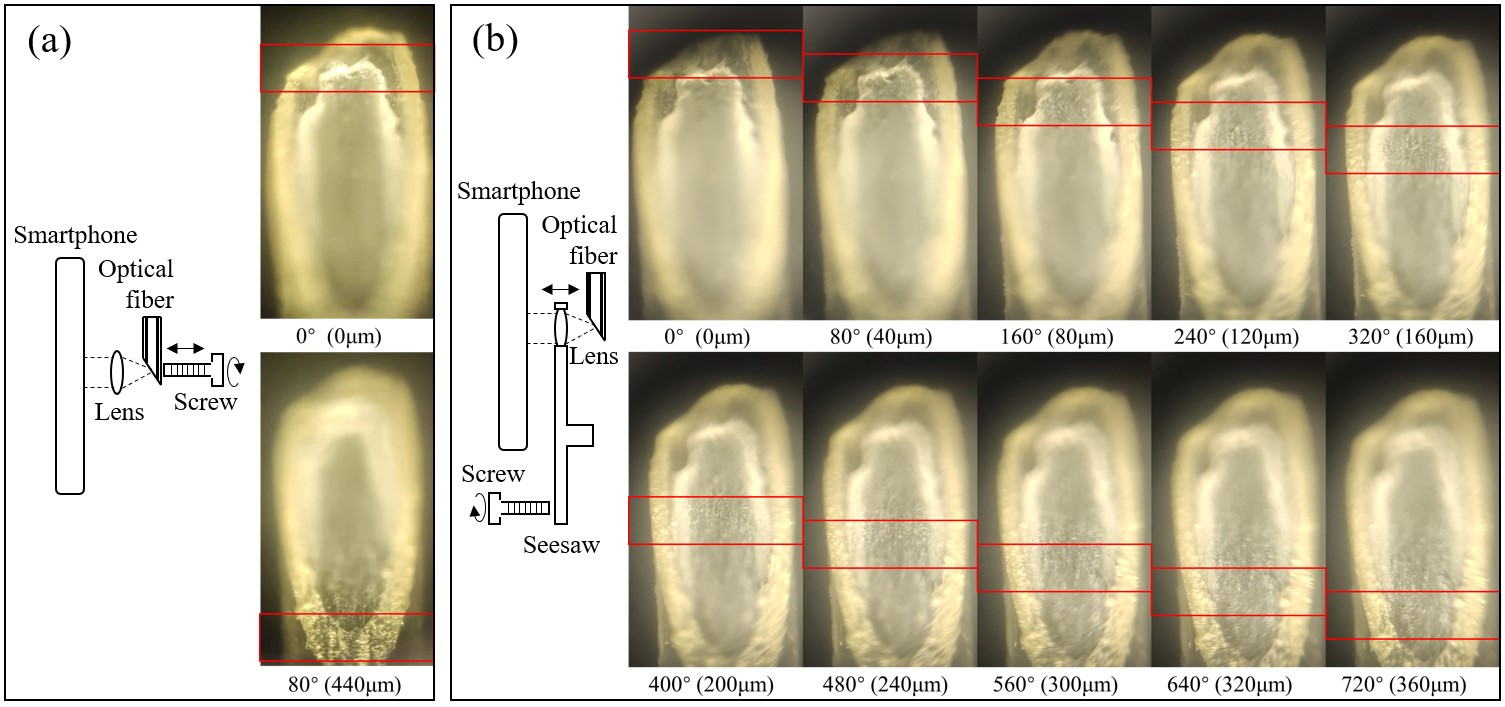}
\caption{Characterizing the focus tunability of our method by imaging the 30-degree facet of a silica optical fiber. We compare the focus tuning by (a) directly changing the position of the fiber via the 3D-printed screw spike and (b) the seesaw-like structure.}
\label{fig7}
\end{figure}
\indent We characterize the focus tunability of our method by imaging the 30-degree facet of a silica optical fiber as illustrated in Fig.~\ref{fig7}. We employ a silica optical fiber with a core/cladding/coating diameter of 400/440/625 $\mu$m (MM-S400/440-12A, Nufern, USA) and polished its end facet into a 30-degree slope, which gives us the vertical features within the FOV. We compare the focus tuning by directly changing the position of the fiber via the 3D-printed screw spike [Fig.~\ref{fig7}(a)] and (b) the seesaw-like structure [Fig.~\ref{fig7}(b)]. Leveraging this characterization method, we first achieve the measurement of a depth of focus $d$ of $\sim40$ $\mu$m. It is in accordance with the theoretical prediction in Equation~\ref{eq0} with a numerical aperture $NA$ of $\sim0.12$, which could be deduced from the overall magnification of $\sim6\times$, the capability of our ultra-low-cost smartphone microscope. As demonstrated in Fig.~\ref{fig7}, by manually rotating the 3D-printed screw spike every 80 degrees, the seesaw-like structure can achieve a much finer focus tuning ($\sim11$ times) than that of directly tuning the position of the fiber via the 3D-printed screw spike.\\
\begin{figure}[h!]
    \centering
    \includegraphics[width=0.8\linewidth]{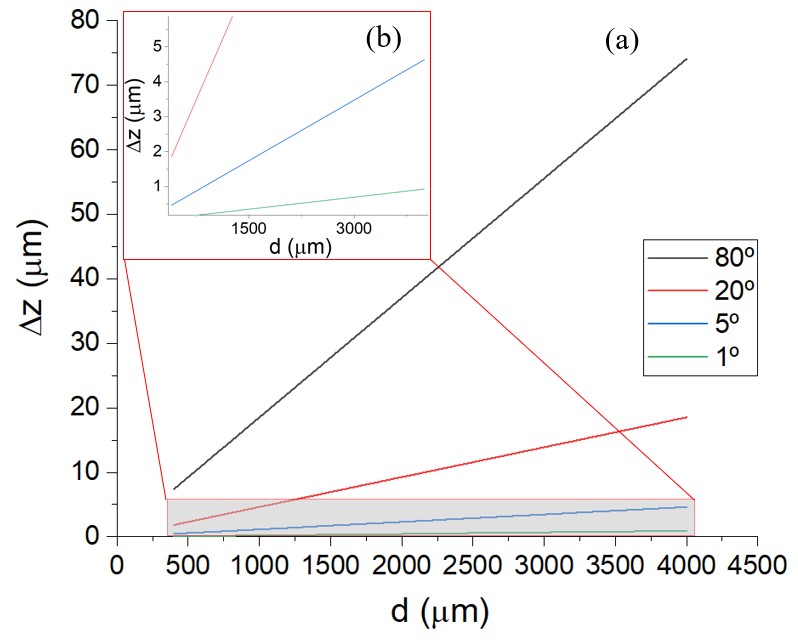}
    \caption{Tuning resolution $\Delta z$ as functions of the thread pitch of 3D-printed screw $d$ and the minimal rotation angle $a$, with the displacement ratio of the force and distal ends of the seesaw-like structure $r$ of 11.}
    \label{fig:acc}
\end{figure}
\indent For such a focus tuning method, the accuracy $\Delta z$ could be expressed as
\begin{equation}
    \Delta z = \frac{a\cdot d}{2\pi\cdot r},
\end{equation}
where $d$ is the flutes pitch of 3D-printed screw. $a$ is the minimal rotation angle. $r$ is the displacement ratio of the force and distal ends of the seesaw-like structure. With a $r$ of 11 achieved in this work, we plotted the $\Delta z$ as functions of $a$ and $d$ as shown in Fig.~\ref{fig:acc}. We can see the tuning resolution improves as the $a$ and $d$ decrease. sub-micron-scale resolution could be easily achieved with an $a$ of $<1$ degree. In our experiments, we found an $a$ of 5 degrees could be achieved via hand-twisting of the screw, which indicates a focus tuning accuracy of better than 5 $\mu$m. To further improve the minimal rotation angle in a hand-twisting setting, a larger diameter of the screw should be employed. With electrical controls such as a stepping motor, an $a$ of $<1$ degree can be achieved, which leads to sub-micron focus tuning accuracy.
\subsection{Imaging resolution of ultra-low-cost smartphone microscope}
\begin{figure}[h!]
\centering
\includegraphics[width=\linewidth]{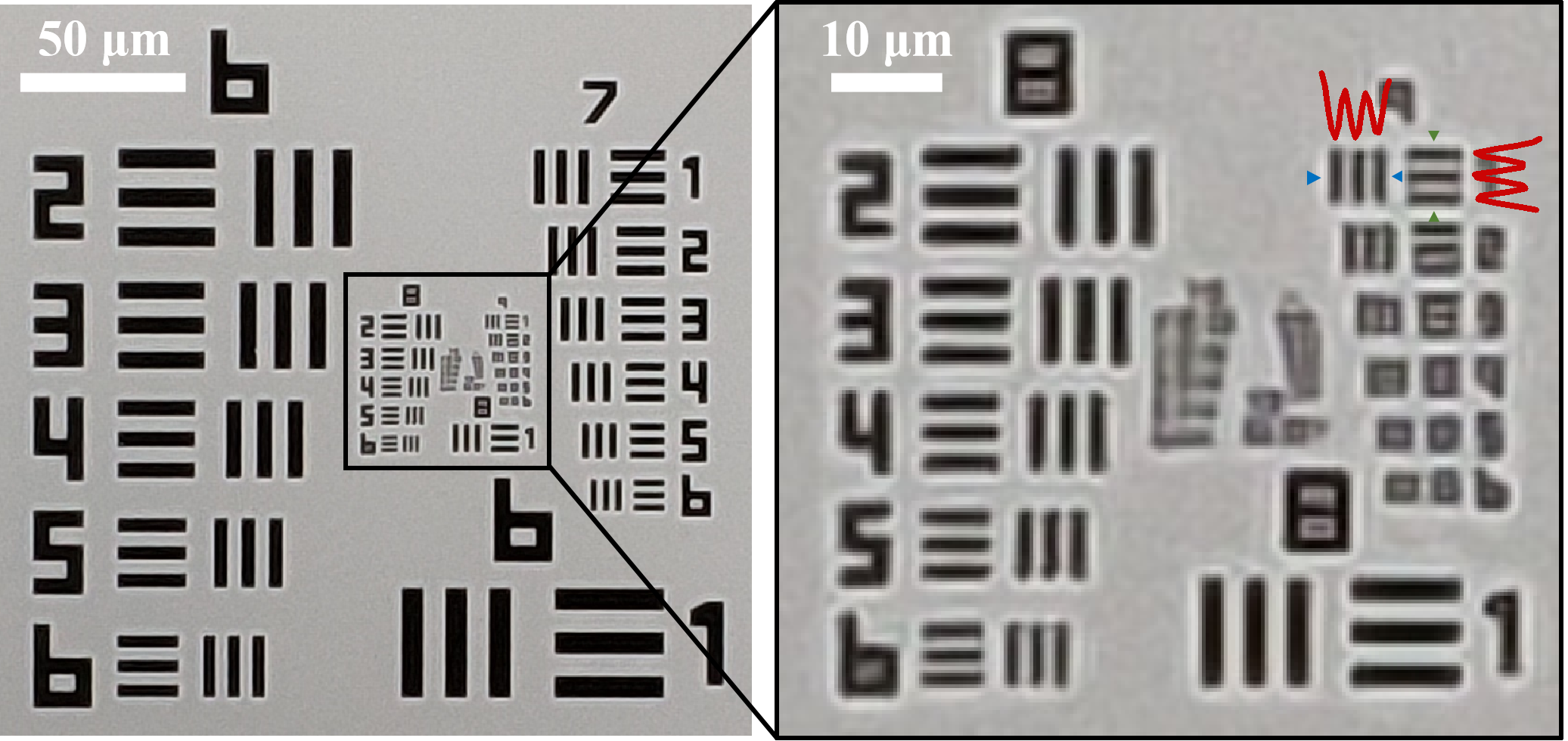}
\caption{Imaging a USAF-1951 resolution target using our ultra-low-cost smartphone microscope. A video of its focus-tuning process is given in Visualization 1 of the supplementary materials.}
\label{fig:res}
\end{figure}
\indent We characterize the imaging performance of our ultra-low-cost smartphone microscope with a USAF-1951 resolution target (HIGHRES-1, Newport, USA). As can be seen on the left side of Fig.~\ref{fig:res}, our smartphone microscope has no vignetting effect along with low aberration. As can be seen on the right side of the figure, our microscope has the ability to resolve to the third element of Group 9 of the USAF-1951 resolution target, which corresponds to a spatial resolution of 0.78 $\mu$m. However, it can be seen that the microscope did not successfully resolve at the end of this element. So we used the first element of Group 9 as a measure of the resolution of our ultra-low-cost smartphone microscope, which corresponds to a spatial resolution of 0.98 $\mu$m. Besides, as mentioned above, the seesaw-like structure has a horizontal displacement during the focus tuning. In the supplementary materials, we give a video of the focus tuning process during the imaging of the USAF-1951 target (Visualization 1), which justifies the negligible influence of the horizontal displacement on the FOV. 
\begin{figure}[h!]
\centering\includegraphics[width=\linewidth]{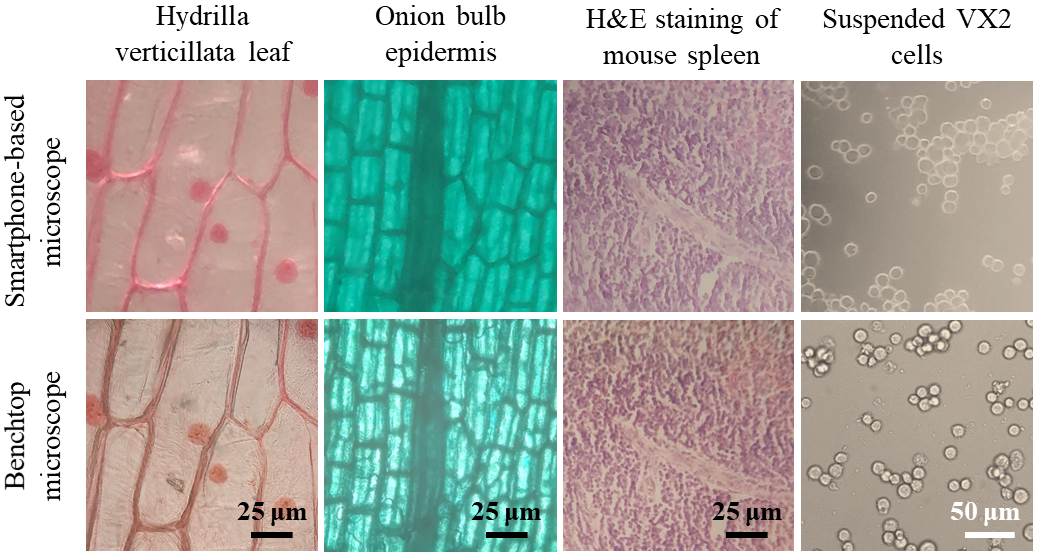}
\caption{Imaging performance comparison between our smartphone-based microscope and a $20\times$ benchtop microscope using different biomedical specimens.}
\label{fig8}
\end{figure}
\begin{figure}[h!]
    \centering
    \includegraphics[width=\linewidth]{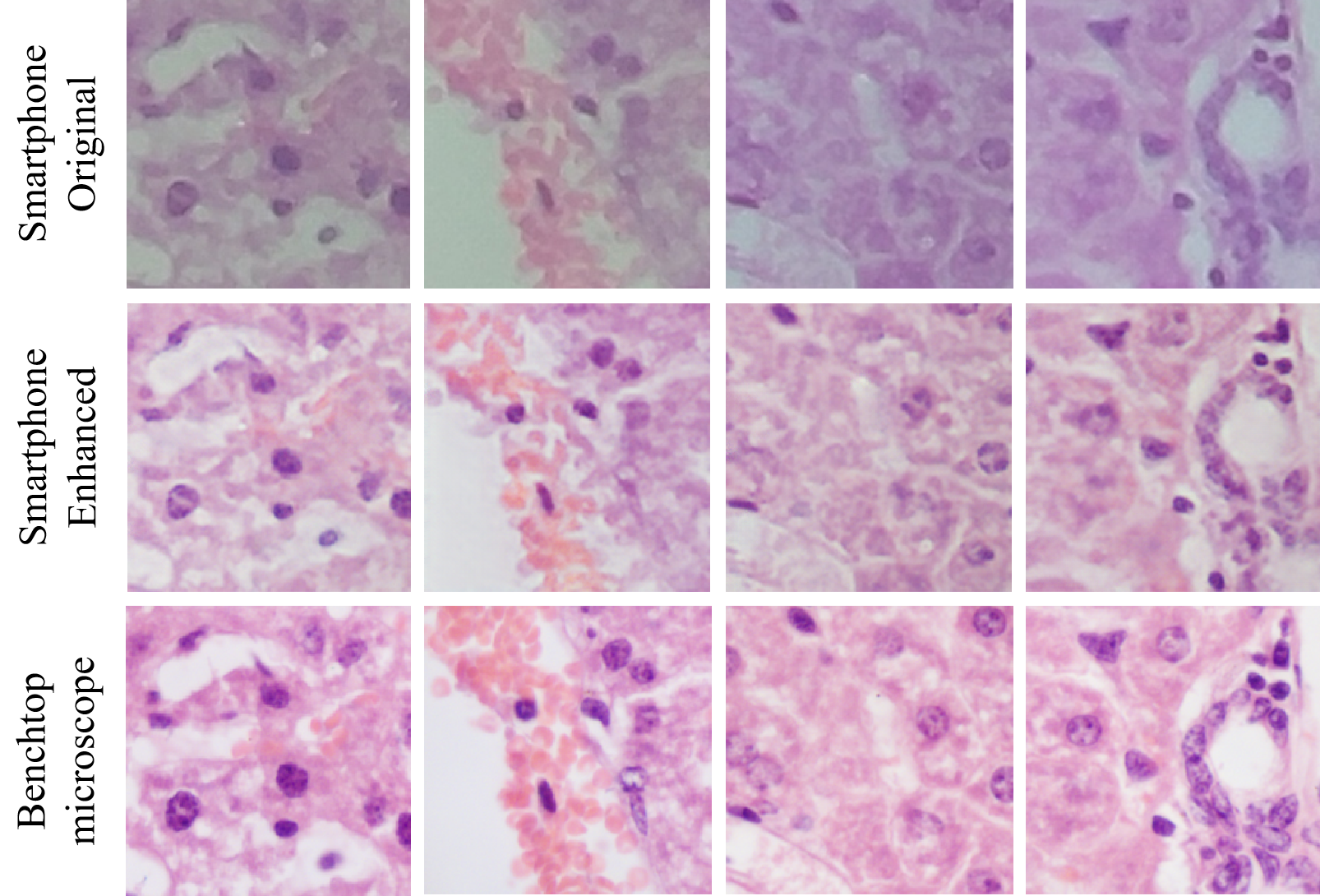}
    \caption{Examples of enhanced smartphone microscopic images using unpaired deep learning. We compare them with the original ones and their benchtop counterparts.}
    \label{fig:enhanced}
\end{figure}
\subsection{Comparision of our smartphone microscope with a benchtop microscope}
Figure~\ref{fig8} demonstrates the imaging performance comparison between our ultra-low-cost smartphone microscope and a $20\times$ benchtop microscope (Axioplan 2, Zeiss, Germany) using different biomedical specimens, including hydrilla verticillata leaf, onion bulb epidermis, hemotoxylin and eosin (H$\&$E) staining of mouse spleen, and suspended VX2 cells. We can see that the resolving capability of biological tissues and cells using our smartphone microscope is close to that of using the benchtop microscope. However, the imaging contrast of the smartphone microscope is inferior to that of the benchtop microscope, which may be attributed to our low-cost illumination setup. In comparison, the benchtop microscope uses the K{\"o}hler illumination, which enables better imaging contrast.
\subsection{Enhanced smartphone microscopy using unpaired deep learning}
To address the above problem of imaging contrast, we use the developed unpaired deep learning method to enhance the imaging quality. Here we take the stained slices of a mouse liver as an example. The results are shown in Fig.~\ref{fig:enhanced}. It can be seen that the contrast of the smartphone microscopic images has been significantly improved after using our trained unpaired deep learning model. Compared with the imaging results of the benchtop microscope in the bottom row, the image enhancement algorithm does not bring significant artifacts and structural changes, which indicates its application potential.
\begin{figure}[h!]
    \centering
    \includegraphics[width=\linewidth]{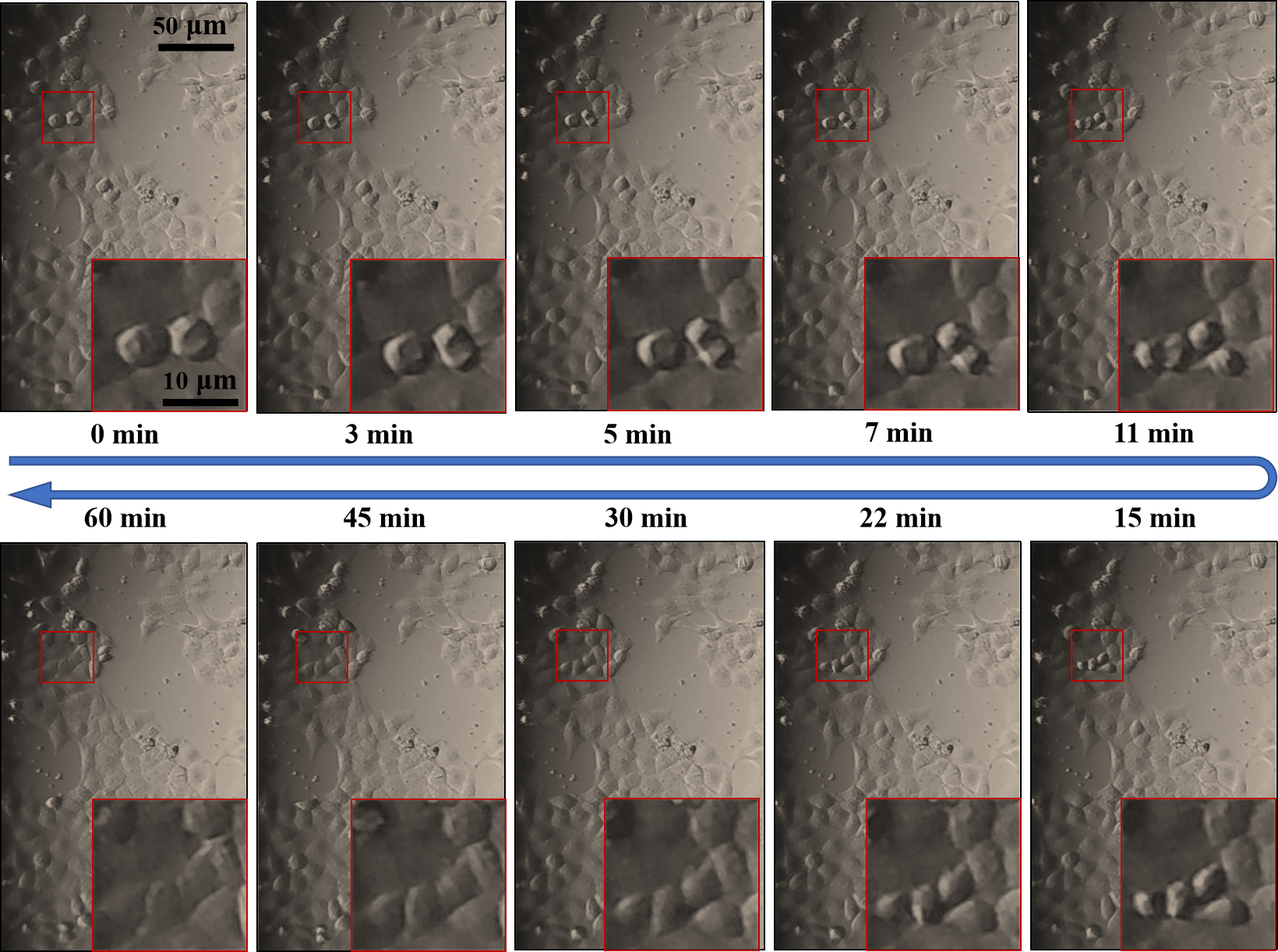}
    \caption{Application of our ultra-low-cost smartphone microscope in the monitoring of VX2 tumor cell culture.}
    \label{fig:vx2}
\end{figure}
\subsection{Application in the monitoring of VX2 tumor cell culture}
We apply our ultra-low-cost smartphone microscope in the monitoring of VX2 tumor cell culture as demonstrated in Fig.~\ref{fig:vx2}. The VX2 is a well-known animal tumor model in interventional radiology \cite{kidd1940transplantable}. It is an anaplastic squamous cell carcinoma caused by infection with the rabbit Shope papillomavirus. This tumor can be transplanted serially from one animal to another by means of xenotransplantation and can grow in any organ or grafted tissue. A notable feature of VX2 is that it does not require genetically modified or immunocompromised subjects but can be transplanted into immunocompetent animals \cite{pascale2022rabbit}. We observed morphological differences in VX2 tumor cells across multiple focal planes and documented the biological process of cell division \textit{in situ} over a period of 1 hour, demonstrating the ability of our ultra-low-cost smartphone microscope to have real-time microscopic monitoring of closed cell culture processes.
\subsection{Fabrication cost of our smartphone microscope}
\begin{table}[h!]
    \centering
        \caption{Bill of materials used to fabricate our smartphone microscope.}
    \begin{tabular}{ccc}
    \hline
        Component & Source & Price (USD)\\
        \hline
         3D-printed enclosure & WeNext Technology Co., Ltd. & 1.7\\
         Lens taken from an iPhone 6s front camera & taobao.com & 2\\
         \hline
         \multicolumn{2}{c}{Total fabrication cost} & 3.7 \\
         \hline
    \end{tabular}
    \label{tab:cost}
\end{table}
Table~\ref{tab:cost} lists the bill of materials used to fabricate our smartphone microscope. It only includes two components. The first component is the 3D-printed enclosure, which includes the seesaw-like structure, the screw spike, the objective lens hold, and the connection part with the smartphone (as shown in Fig.~\ref{fig5}). We customized this component from WeNext Technology Co., Ltd. and the cost is 1.7 USD. The second component is the objective lens taken from an iPhone 6s front camera. It was purchased from taobao.com and the cost is 2 USD. So the total cost to fabricate our smartphone microscope is 3.7 USD.
\section{Discussion}
For previous works to lower the costs of smartphone microscopes, the capability of focus tuning and the imaging quality are sacrificed. In this work, we propose ultra-low-cost solutions for optical focus tuning and image quality degradation.\\
\indent In the field of optical engineering, there are usually two different pathways to handle the out-the-focus issue \cite{bass2010handbook}. One is to extend the depth of field via various techniques \cite{zalevsky2010extended}, such as wavefront shaping \cite{jin2020deep}, multiple aperture synthesis \cite{bo2017depth}, and spectral multiplexing \cite{abrahamsson2013fast}. However, wavefront shaping methods, which employ axicon lens or pupil filter (phase or amplitude mask) for converting Gaussian beam to Bessel beam, suffer a major loss of spatial frequency components in the coherence transfer function thus degrading image quality \cite{bo2017depth}. Synthesis- or multiplexing-based methods, on the other hand, significantly increase the complexity of optics, which is difficult to implement on smartphone-based devices. Besides, the extending depth of focus methods mentioned above all greatly boost the cost of fabrication, thus are suboptimal for the design of smartphone-based biomedical instrumentation.\\
\indent The other pathway to address the out-of-focus issue is to adjust the position of focus. Actually, there are built-in tuning components inside high-end smartphones. However, we found it had a very limited tuning range that is not enough for compensating the defocusing caused by the inaccurate fabrication of 3D printing. Electrical- or mechanical-driven liquid lenses \cite{song2020smartphone,yang2017handheld} could be a convenient approach for large-range focus tuning, but they also have the shortcomings of image aberration and increased cost.\\
\indent The proposed method merely leverages the flexibility of 3D printing materials via a seesaw-like structure, so it does not add cost and complexity to smartphone microscopy. The fine-tuning of optical focus could be realized by a single 3D-printed screw as demonstrated and discussed in Section 3.2.\\
\indent On the other hand, previous smartphone microscope works used the objective lens and external LED illumination to keep the imaging quality \cite{sun2018low,trofymchuk2021addressable}, how such approaches significantly increase the cost of fabrication. Lee and Yang employed ambient sunlight for illumination and the user’s hand motion for angular scanning \cite{lee2014smartphone}. This approach leads to low cost but brings instability to the imaging process. Orth \textit{et al.} used the onboard camera flashlight for illumination \cite{orth2018dual}. To realize transmitted illumination, they employed an additional mirror to reflect the flashlight. During the microscopic imaging, they turned the mode of the built-in flashlight from flash-light to fill-light. Our illumination strategy is similar to this, except that we use reflected illumination and thus do not need an extra mirror. Both of us employ the camera lenses of smartphones. However, such methods lower the imaging quality. \\
\indent To address this issue, we introduce a deep-learning-based image enhancement method. Compared with the previous works that used deep learning in smartphone microscopes \cite{rivenson2018deep,kim2023portable}, we leverage the cycle-consistency generative adversarial network architecture \cite{zhu2017unpaired} to establish the mapping from smartphone microscope images to benchtop microscope images, which do not need the paired field of view data for training.\\
\indent Although we have demonstrated the focus tunability and imaging quality of our ultra-low-cost smartphone microscopy. Its intrinsic limitations should be aware before utilization. (1) Besides the axial displacement we used for the focus tuning, there is a transverse displacement exists, which is $\sim7.3$ times smaller than the axial displacement. We found it had no visible influence on the performance of our smartphone microscope. But for higher magnification, this alteration of FOV during focus tuning may raise issues. (2) By longitudinally monitoring the resetting position of the seesaw-like structure for 150 times, We have preliminarily verified the repeatability of our focus tuning method. However, based on the fatigue characteristics of 3D printing materials \cite{yap2020review}, its performance may still degrade after long-term usage. (3) For different imaging object types, our deep-learning-based image enhancement method needs to recollect and retrain the model. Although the introduction of unpaired learning has made these works easy, it is still tedious for the smartphone microscopes that are used in multiple scenarios. 
\section{Conclusions}
In response to the application needs of home health detection after the epidemic, here we proposed a solution for ultra-low-cost smartphone microscopy. Utilizing the flexural properties of the 3D printed material itself, we designed a seesaw-like mechanical structure that is able to convert large displacements on one side into small displacements on the other side (reduced to ~9.1\%), which in turn achieves a focusing accuracy of $\sim5$ $\mu$m, which is 40 times higher than the machining accuracy of the 3D printing itself. The phone's own flash was used as the microscopic imaging illumination, and an off-the-shelf phone camera lens was used as the objective lens. To compensate for the degradation of image quality due to this relatively simple microscope configuration, we developed a deep learning-based image enhancement algorithm that can effectively improve the image quality. We applied the device to mouse hepatocyte-stained slice imaging and VX2 tumor cell culture observation. Our method is able to keep the full cost of the device, except for the smartphone, below 4 USD, which will help the application and promotion of smartphone biomedical detection technologies.

\begin{backmatter}

\bmsection{Funding}
This work was partially supported by National Natural Science Foundation of China under Grant No. 62105198.
\bmsection{Acknowledgments}
We would like to thank the editors and reviewers for their time and effort in helping us improve this manuscript.
\bmsection{Disclosures}
The authors declare no conflicts of interest.
\end{backmatter}

%%%%%%%%%% If using BibTeX:
\bibliography{sample}

\begin{thebibliography}{10}
\newcommand{\enquote}[1]{``#1''}

\bibitem{banik2021recent}
S.~Banik, S.~K. Melanthota, Arbaaz, J.~M. Vaz, V.~M. Kadambalithaya, I.~Hussain, S.~Dutta, and N.~Mazumder, \enquote{Recent trends in smartphone-based detection for biomedical applications: a review,} {\protect\JournalTitle{Analytical and Bioanalytical Chemistry}} \textbf{413}, 2389--2406 (2021).

\bibitem{hussain2021smartphone}
I.~Hussain and A.~K. Bowden, \enquote{Smartphone-based optical spectroscopic platforms for biomedical applications: a review,} {\protect\JournalTitle{Biomedical optics express}} \textbf{12}, 1974--1998 (2021).

\bibitem{hunt2021smartphone}
B.~Hunt, A.~J. Ruiz, and B.~W. Pogue, \enquote{Smartphone-based imaging systems for medical applications: a critical review,} {\protect\JournalTitle{Journal of Biomedical Optics}} \textbf{26}, 040902--040902 (2021).

\bibitem{ravindran2021smartphone}
S.~Ravindran, \enquote{Smartphone science: apps test and track infectious diseases.} {\protect\JournalTitle{Nature}} \textbf{593}, 302--303 (2021).

\bibitem{fozouni2021amplification}
P.~Fozouni, S.~Son, M.~D. de~Le{\'o}n~Derby, G.~J. Knott, C.~N. Gray, M.~V. D’Ambrosio, C.~Zhao, N.~A. Switz, G.~R. Kumar, S.~I. Stephens \emph{et~al.}, \enquote{Amplification-free detection of sars-cov-2 with crispr-cas13a and mobile phone microscopy,} {\protect\JournalTitle{Cell}} \textbf{184}, 323--333 (2021).

\bibitem{ganguli2020rapid}
A.~Ganguli, A.~Mostafa, J.~Berger, M.~Y. Aydin, F.~Sun, S.~A.~S. de~Ramirez, E.~Valera, B.~T. Cunningham, W.~P. King, and R.~Bashir, \enquote{Rapid isothermal amplification and portable detection system for sars-cov-2,} {\protect\JournalTitle{Proceedings of the National Academy of Sciences}} \textbf{117}, 22727--22735 (2020).

\bibitem{jin2020deep}
L.~Jin, Y.~Tang, Y.~Wu, J.~B. Coole, M.~T. Tan, X.~Zhao, H.~Badaoui, J.~T. Robinson, M.~D. Williams, A.~M. Gillenwater \emph{et~al.}, \enquote{Deep learning extended depth-of-field microscope for fast and slide-free histology,} {\protect\JournalTitle{Proceedings of the National Academy of Sciences}} \textbf{117}, 33051--33060 (2020).

\bibitem{sun2018low}
D.~Sun and T.~Y. Hu, \enquote{A low cost mobile phone dark-field microscope for nanoparticle-based quantitative studies,} {\protect\JournalTitle{Biosensors and Bioelectronics}} \textbf{99}, 513--518 (2018).

\bibitem{knowlton20173d}
S.~Knowlton, A.~Joshi, P.~Syrrist, A.~F. Coskun, and S.~Tasoglu, \enquote{3d-printed smartphone-based point of care tool for fluorescence-and magnetophoresis-based cytometry,} {\protect\JournalTitle{Lab on a Chip}} \textbf{17}, 2839--2851 (2017).

\bibitem{orth2018dual}
A.~Orth, E.~Wilson, J.~Thompson, and B.~Gibson, \enquote{A dual-mode mobile phone microscope using the onboard camera flash and ambient light,} {\protect\JournalTitle{Scientific reports}} \textbf{8}, 3298 (2018).

\bibitem{liu2021pocket}
Y.~Liu, A.~M. Rollins, R.~M. Levenson, F.~Fereidouni, and M.~W. Jenkins, \enquote{Pocket muse: an affordable, versatile and high-performance fluorescence microscope using a smartphone,} {\protect\JournalTitle{Communications biology}} \textbf{4}, 334 (2021).

\bibitem{zhu2011optofluidic}
H.~Zhu, S.~Mavandadi, A.~F. Coskun, O.~Yaglidere, and A.~Ozcan, \enquote{Optofluidic fluorescent imaging cytometry on a cell phone,} {\protect\JournalTitle{Analytical chemistry}} \textbf{83}, 6641--6647 (2011).

\bibitem{ganguli2017hands}
A.~Ganguli, A.~Ornob, H.~Yu, G.~Damhorst, W.~Chen, F.~Sun, A.~Bhuiya, B.~Cunningham, and R.~Bashir, \enquote{Hands-free smartphone-based diagnostics for simultaneous detection of zika, chikungunya, and dengue at point-of-care,} {\protect\JournalTitle{Biomedical microdevices}} \textbf{19}, 1--13 (2017).

\bibitem{koydemir2015rapid}
H.~C. Koydemir, Z.~Gorocs, D.~Tseng, B.~Cortazar, S.~Feng, R.~Y.~L. Chan, J.~Burbano, E.~McLeod, and A.~Ozcan, \enquote{Rapid imaging, detection and quantification of giardia lamblia cysts using mobile-phone based fluorescent microscopy and machine learning,} {\protect\JournalTitle{Lab on a Chip}} \textbf{15}, 1284--1293 (2015).

\bibitem{trofymchuk2021addressable}
K.~Trofymchuk, V.~Glembockyte, L.~Grabenhorst, F.~Steiner, C.~Vietz, C.~Close, M.~Pfeiffer, L.~Richter, M.~L. Sch{\"u}tte, F.~Selbach \emph{et~al.}, \enquote{Addressable nanoantennas with cleared hotspots for single-molecule detection on a portable smartphone microscope,} {\protect\JournalTitle{Nature communications}} \textbf{12}, 1--8 (2021).

\bibitem{lee2021smartphone}
K.~C. Lee, K.~Lee, J.~Jung, S.~H. Lee, D.~Kim, and S.~A. Lee, \enquote{A smartphone-based fourier ptychographic microscope using the display screen for illumination,} {\protect\JournalTitle{ACS Photonics}} \textbf{8}, 1307--1315 (2021).

\bibitem{lee2014fabricating}
W.~Lee, A.~Upadhya, P.~Reece, and T.~G. Phan, \enquote{Fabricating low cost and high performance elastomer lenses using hanging droplets,} {\protect\JournalTitle{Biomedical optics express}} \textbf{5}, 1626--1635 (2014).

\bibitem{szydlowski2020cell}
N.~A. Szydlowski, H.~Jing, M.~Alqashmi, and Y.~S. Hu, \enquote{Cell phone digital microscopy using an oil droplet,} {\protect\JournalTitle{Biomedical Optics Express}} \textbf{11}, 2328--2338 (2020).

\bibitem{dai2019colour}
B.~Dai, Z.~Jiao, L.~Zheng, H.~Bachman, Y.~Fu, X.~Wan, Y.~Zhang, Y.~Huang, X.~Han, C.~Zhao \emph{et~al.}, \enquote{Colour compound lenses for a portable fluorescence microscope,} {\protect\JournalTitle{Light: Science \& Applications}} \textbf{8}, 1--13 (2019).

\bibitem{satzuma}
\url{https://www.satzuma.com/product-page/smartphone-microscope}.

\bibitem{apexel}
\url{https://www.apexeloptic.com/product/200x-phone-microscope-lens/}.

\bibitem{song2020smartphone}
C.~Song, Y.~Yang, X.~Tu, Z.~Chen, J.~Gong, and C.~Lin, \enquote{A smartphone-based fluorescence microscope with hydraulically driven optofluidic lens for quantification of glucose,} {\protect\JournalTitle{IEEE Sensors Journal}} \textbf{21}, 1229--1235 (2020).

\bibitem{zhu2017unpaired}
J.-Y. Zhu, T.~Park, P.~Isola, and A.~A. Efros, \enquote{Unpaired image-to-image translation using cycle-consistent adversarial networks,} in \emph{Proceedings of the IEEE international conference on computer vision,}  (2017), pp. 2223--2232.

\bibitem{chondros2010archimedes}
T.~G. Chondros, \enquote{Archimedes life works and machines,} {\protect\JournalTitle{Mechanism and Machine Theory}} \textbf{45}, 1766--1775 (2010).

\bibitem{gross2016mechanics}
D.~Gross, W.~Ehlers, P.~Wriggers, J.~Schr{\"o}der, and R.~M{\"u}ller, \emph{Mechanics of Materials-Formulas and Problems} (Springer, 2016).

\bibitem{lu2020universal}
Y.~Lu and J.~Lu, \enquote{A universal approximation theorem of deep neural networks for expressing probability distributions,} {\protect\JournalTitle{Advances in neural information processing systems}} \textbf{33}, 3094--3105 (2020).

\bibitem{rivenson2018deep}
Y.~Rivenson, H.~Ceylan~Koydemir, H.~Wang, Z.~Wei, Z.~Ren, H.~G{\"u}nayd{\i}n, Y.~Zhang, Z.~Gorocs, K.~Liang, D.~Tseng \emph{et~al.}, \enquote{Deep learning enhanced mobile-phone microscopy,} {\protect\JournalTitle{Acs Photonics}} \textbf{5}, 2354--2364 (2018).

\bibitem{kim2023portable}
K.~Kim and W.~G. Lee, \enquote{Portable, automated and deep-learning-enabled microscopy for smartphone-tethered optical platform towards remote homecare diagnostics: A review,} {\protect\JournalTitle{Small Methods}} \textbf{7}, 2200979 (2023).

\bibitem{switz2014low}
N.~A. Switz, M.~V. D'Ambrosio, and D.~A. Fletcher, \enquote{Low-cost mobile phone microscopy with a reversed mobile phone camera lens,} {\protect\JournalTitle{PloS one}} \textbf{9}, e95330 (2014).

\bibitem{pytorch}
\url{https://github.com/junyanz/pytorch-CycleGAN-and-pix2pix}.

\bibitem{kidd1940transplantable}
J.~G. Kidd and P.~Rous, \enquote{A transplantable rabbit carcinoma originating in a virus-induced papilloma and containing the virus in masked or altered form,} {\protect\JournalTitle{The Journal of experimental medicine}} \textbf{71}, 813--838 (1940).

\bibitem{pascale2022rabbit}
F.~Pascale, J.-P. Pelage, M.~Wassef, S.~H. Ghegediban, J.-P. Saint-Maurice, T.~De~Baere, A.~Denys, R.~Duran, F.~Deschamps, O.~Pellerin \emph{et~al.}, \enquote{Rabbit vx2 liver tumor model: a review of clinical, biology, histology, and tumor microenvironment characteristics,} {\protect\JournalTitle{Frontiers in Oncology}} \textbf{12}, 871829 (2022).

\bibitem{bass2010handbook}
M.~Bass, \emph{Handbook of Optics} (McGraw-Hill Education, 2010).

\bibitem{zalevsky2010extended}
Z.~Zalevsky, \enquote{Extended depth of focus imaging: a review,} {\protect\JournalTitle{Spie Reviews}} \textbf{1}, 018001 (2010).

\bibitem{bo2017depth}
E.~Bo, Y.~Luo, S.~Chen, X.~Liu, N.~Wang, X.~Ge, X.~Wang, S.~Chen, S.~Chen, J.~Li \emph{et~al.}, \enquote{Depth-of-focus extension in optical coherence tomography via multiple aperture synthesis,} {\protect\JournalTitle{Optica}} \textbf{4}, 701--706 (2017).

\bibitem{abrahamsson2013fast}
S.~Abrahamsson, J.~Chen, B.~Hajj, S.~Stallinga, A.~Y. Katsov, J.~Wisniewski, G.~Mizuguchi, P.~Soule, F.~Mueller, C.~D. Darzacq \emph{et~al.}, \enquote{Fast multicolor 3d imaging using aberration-corrected multifocus microscopy,} {\protect\JournalTitle{Nature methods}} \textbf{10}, 60--63 (2013).

\bibitem{yang2017handheld}
J.~Yang, L.~Liu, J.~P. Campbell, D.~Huang, and G.~Liu, \enquote{Handheld optical coherence tomography angiography,} {\protect\JournalTitle{Biomedical optics express}} \textbf{8}, 2287--2300 (2017).

\bibitem{lee2014smartphone}
S.~A. Lee and C.~Yang, \enquote{A smartphone-based chip-scale microscope using ambient illumination,} {\protect\JournalTitle{Lab on a Chip}} \textbf{14}, 3056--3063 (2014).

\bibitem{yap2020review}
Y.~L. Yap, S.~L. Sing, and W.~Y. Yeong, \enquote{A review of 3d printing processes and materials for soft robotics,} {\protect\JournalTitle{Rapid Prototyping Journal}}  (2020).

\end{thebibliography}

%%%%%%%%%% If preparing manually:
% \newcommand{\enquote}[1]{``#1''}

% \bibitem{Zhang:14}
% Y.~Zhang, S.~Qiao, L.~Sun, Q.~W. Shi, W.~Huang, L.~Li, and Z.~Yang,
%   \enquote{Photoinduced active terahertz metamaterials with nanostructured
%   vanadium dioxide film deposited by sol-gel method,}
%   {\protect\JournalTitle{Optics Express}} \textbf{22}, 11070--11078 (2014).

% {Optical Society}, \enquote{{OSA Publishing},}
%   \url{http://www.osapublishing.org}.

% \bibitem{FORSTER2007}
% P.~Forster, V.~Ramaswamy, P.~Artaxo, T.~Bernsten, R.~Betts, D.~Fahey,
%   J.~Haywood, J.~Lean, D.~Lowe, G.~Myhre, J.~Nganga, R.~Prinn, G.~Raga,
%   M.~Schulz, and R.~V. Dorland, \enquote{Changes in atmospheric consituents and
%   in radiative forcing,} in \enquote{Climate Change 2007: The Physical Science
%   Basis. Contribution of Working Group 1 to the Fourth assesment report of
%   Intergovernmental Panel on Climate Change,}  S.~Solomon, D.~Qin, M.~Manning,
%   Z.~Chen, M.~Marquis, K.~B. Averyt, M.~Tignor, and H.~L. Miler, eds.
%   (Cambridge University Press, 2007).

% \end{thebibliography}

\end{document}